\documentclass[aps,nofootinbib,showpacs,preprint,superscriptaddress]{revtex4}
\usepackage{graphicx}
\usepackage{amsthm}
\usepackage{amsmath}
\usepackage{amssymb}
\usepackage{mathrsfs}
\usepackage{color}
\usepackage{graphics}
\usepackage{ccaption}
\def\slashchar#1{\setbox0=\hbox{$#1$}  
   \dimen0=\wd0     
   \setbox1=\hbox{/} \dimen1=\wd1  
   \ifdim\dimen0>\dimen1   
      \rlap{\hbox to \dimen0{\hfil/\hfil}} 
      #1     
   \else     
      \rlap{\hbox to \dimen1{\hfil$#1$\hfil}} 
      /      
   \fi}

\makeatletter
\def\overbracket#1{\mathop{\vbox{\ialign{##\crcr\noalign{\kern3\p@}
\downbracketfill\crcr\noalign{\kern3\p@\nointerlineskip}
$\hfil\displaystyle{#1}\hfil$\crcr}}}\limits}
\def\underbracket#1{\mathop{\vtop{\ialign{##\crcr
$\hfil\displaystyle{#1}\hfil$\crcr\noalign{\kern3\p@\nointerlineskip}
\upbracketfill\crcr\noalign{\kern3\p@}}}}\limits}
\def\upbracketfill{$\m@th\makesm@sh{\llap{\vrule\@height3\p@\@width.7\p@}}%
\leaders\vrule\@height.7\p@\hfill
\makesm@sh{\rlap{\vrule\@height3\p@\@width.7\p@}}$}
\def\downbracketfill{$\m@th
\makesm@sh{\llap{\vrule\@height.7\p@\@depth2.3\p@\@width.7\p@}}%
\leaders\vrule\@height.7\p@\hfill
\makesm@sh{\rlap{\vrule\@height.7\p@\@depth2.3\p@\@width.7\p@}}$}
\makeatother

\begin{document}




\title{
Axial anomaly and the interplay of quark loops with pseudoscalar 
and vector mesons in the $\gamma^{*}\to\pi^{+}\pi^{0}\pi^{-}$ process
}

\author{Sanjin Beni\'{c}}
\affiliation{\footnotesize Department of Physics, Faculty of Science,
        Zagreb University, Bijeni\v{c}ka c. 32, 10000 Zagreb, Croatia}

\author{Dubravko Klabu\v{c}ar\footnote{Senior associate of Abdus Salam ICTP, 
corresponding author, e-mail: klabucar@phy.hr}}
\affiliation{\footnotesize Department of Physics, Faculty of Science,
        Zagreb University, Bijeni\v{c}ka c. 32, 10000 Zagreb, Croatia}

\date{\today}

\begin{abstract}
Motivated by the ongoing measurements of the Primakoff process 
$\pi^{-}\gamma^{*}\to\pi^{-}\pi^{0}$ by COMPASS collaboration at CERN, the transition
form factor for the canonical anomalous process $\gamma^{*}\to\pi^{+}\pi^{0}\pi^{-}$
is calculated in a constituent quark loop model.
The simplest contribution to this process is the quark ``box" amplitude.
In the present paper we also explicitly include the vector meson degrees of freedom,
 i.e., the $\rho$ and the $\omega$, thus giving rise to 
additional, resonant contributions. We find that in order to
  satisfy the axial anomaly
result, a further subtraction in the resonant part is needed. The
results are then compared with the vector meson dominance model as well 
as the Dyson--Schwinger calculations, the chiral perturbation theory result,
and the available data. 
\end{abstract}

\maketitle

\section{Introduction}
\label{intro}

\begin{figure}[b!]
\begin{center}
\includegraphics[scale=0.8]{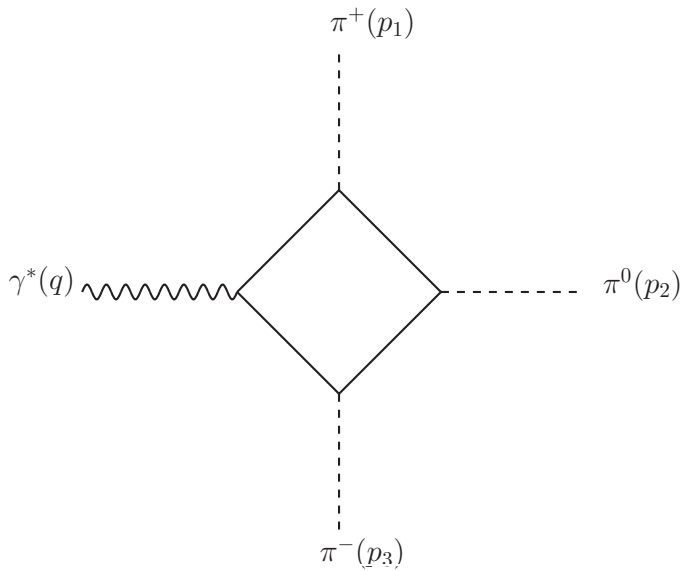}
\caption{One of the box diagrams for the process $\gamma^{*}\to \pi^+ \pi^0 \pi^-$. There are six different contributing graphs, obtained from Fig. \ref{figbox} by the permutations of the vertices of the three different pions. The position of the $u$ and $d$ quark flavors on the internal lines, as well as $Q_u$ or $Q_d$ quark charges in the quark-photon vertex, varies from graph to graph, depending on the position of the quark-pion vertices. The physical pion fields are $\pi^\pm=(\pi_1\mp i \pi_2)/\sqrt{2}$ and $\pi^0\equiv\pi_3$. Thus, in Eq. (\ref{1Leff}) one has $\pi =\sqrt{2}(\pi^+t_+ + \pi^- t_-) + \pi^0 t_3$ where $t_{\pm}=t_1 \pm it_2$ (see text).}
\label{figbox}
\end{center}
\end{figure}

The electromagnetic processes influenced by the Abelian axial anomaly \cite{Adler:1969gk, Bell:1969ts} are of considerable
theoretical interest. Among them are the transitions of the type $\gamma^*(q) \to P^+(p_1) P^0(p_2) P^-(p_3)$, where
$\gamma^*$ denotes a, generally, virtual ($q^2\neq 0$) photon $\gamma$, $P^\pm$ stands for a charged and $P^0$ for a neutral meson from the pseudoscalar nonet, 
up to the strangeness conservation (so that $P^\pm = \pi^\pm, K^\pm$ and 
$P^0 = \pi^0, \eta, \eta'$). 
These processes are supposedly influenced by the, colloquially called, ``box" axial anomaly, since on the microscopic level, the three pseudoscalar ($P$) mesons would couple to the photon through a four-vertex quark loop, like in Fig. \ref{figbox}. 

In the chiral limit (where ${m_\pi=0}$) and the soft--point limit  (of vanishing 4-momenta of external particles, $p_j=0=q$), 
which is a reasonably realistic approximation at low energies at least for the lightest pseudoscalars -- the pions, the anomaly analysis predicts 
\cite{Adler:1971nq,Terent'ev:1971kt,Terentev:1974vr,Aviv:1971hq} 
that the theoretical amplitude is exactly
\begin{equation}
A_{\gamma}^{3\pi}\equiv\lim_{m_{\pi}\to 0}F_\gamma^{3\pi}(p_1=0,p_2=0,p_3=0) = \frac{e \, N_c}{12\pi^2 \, f_\pi^3}  \, ,
\label{boxA}
\end{equation}
where $e$ is the proton charge, $N_c$ the number of quark colors, and the pion decay constant $f_\pi = (92.42 \pm 0.33)$ MeV, whereby 
$A_{\gamma}^{3\pi} = (9.72 \pm 0.09)\, {\rm GeV}^{-3}$.

On the other hand, the experimental knowledge of the processes that should be influenced by the ``box anomaly" is not at all satisfactory, being quite scant. For the $\gamma^*\to \pi^+\pi^0\pi^-$ processes, which should be best approximated by the anomaly prediction (\ref{boxA}) since it involves only the lightest pseudoscalars, there is only one
published 
experimental value for the amplitude at finite momenta $p_j$, i.e., the 
form factor $F_\gamma^{3\pi}(p_1,p_2,p_3)$. 
It was extracted from the cross-section measured \cite{Antipov:1986tp} 
at Serpukhov in the transition $\pi^- \gamma^*\to \pi^0\pi^-$ through the 
Primakoff effect, so that its value
$F_\gamma^{3\pi}(expt) = 12.9 \pm 0.9 \pm 0.5 \, {\rm GeV}^{-3}$ 
really corresponds to the average value of 
the form factor over the momentum range covered by the experiment.
The $\pi^-$ scattering on electrons at CERN SPS yielded the total 
cross section \cite{Amendolia:1985bs} consistent with the Serpukhov value. 
(It is maybe cautious to recall that the both results \cite{Antipov:1986tp,Amendolia:1985bs} are in the strong disagreement with a rarely quoted analysis \cite{Meshcheryakov+al68} of an old measurement \cite{Blokhintseva+al} of this elusive process.) 
In the meantime, one still awaits the analysis of the measurements of this form factor performed at CEBAF \cite{Miskimen94}.

Now, however, there are new hopes of more and better experimental knowledge of such processes, as new high-statistic data on the form factor for the $\pi^- \gamma^* \to \pi^- \pi^0$ transition are expected soon from the COMPASS Primakoff experiments at CERN \cite{Abbon:2007pq,Moinester:1997dm}. (Not only pion, but also kaon beams can be used in these experiments, so that also the reaction $K^- \gamma^* \to K^- \pi^0$ can be studied by the COMPASS collaboration.)

Thus, experiments may finally confirm the relation (\ref{ATAtheorem}) 
between the ``box anomaly" processes and much 
better understood and measured ``triangle anomaly" processes, 
notably the $\pi^0(p)\to \gamma(k_{1})\gamma(k_{2})$ decay into two real photons,
$k_1^2=0=k_2^2$. 
Namely, the pertinent chiral-limit and soft-point amplitudes 
$A_{\pi}^{2\gamma}\equiv\lim_{m_{\pi}\to 0}T_{\pi}(k_{1},k_{2})$ 
and $A_{\gamma}^{3\pi}$ are related 
\cite{Adler:1971nq,Terent'ev:1971kt,Terentev:1974vr,Aviv:1971hq} as
\begin{equation} 
A_{\pi}^{2\gamma} \, = \, {e f_\pi^2} \, A_{\gamma}^{3\pi},
\label{ATAtheorem}
\end{equation}
where, in the notation{\footnote{Except that here, because of 
$F_\gamma^{3\pi}(p_1,p_2,p_3)$ and related function depending on three independent
momenta, it is for brevity not written explicitly, but understood implicitly,
that the scalar functions depend only on the scalar combinations of momenta;
e.g., $T_{\pi}(k_{1},k_{2}) \equiv T_{\pi}(k_{1}^2,k_{2}^2)$ for 
on-shell pions.}} of Ref. \cite{Kekez:1998xr}, $T_{\pi}(k_{1},k_{2})$ 
is the (unnormalized) $\pi^{0}\to \gamma^*\gamma^*$ form factor.

The axial anomaly, which dictates these results, occurs on the level of the quark substructure of hadrons for the quark loops to which 
pseudoscalar mesons are coupled through an odd number of axial (A) vertices, while photons are coupled through vector (V) vertices. 
On the level of effective meson theories where quarks are completely integrated out, the effects of the axial anomaly are encoded in the 
Wess--Zumino--Witten (WZW) Lagrangian term \cite{Wess:1971yu,Witten:1983tw}. Nevertheless, 
thanks to the Veltman-Sutherland theorem{\footnote{Namely, this theorem dictates that in 
the chiral limit, the PVV $\pi^0 \to \gamma\gamma$ amplitude is given exactly by the 
coefficient of the anomaly term (see, e.g., \cite{Pisarski:1997bq,Kekez:2005kx}).},
the anomalous amplitude $A_{\pi}^{2\gamma}$ 
is also obtained successfully through the non-anomalous PVV triangle diagram, where the 
pion is coupled to the fermion loop through the pseudoscalar (P) vertex. 
In the simplest variant, this is achieved through the, basically, ``Steinberger-type" \cite{Steinberger:1949wx} calculation supplemented by the quark--level Goldberger--Treiman (GT) relation $g_{\pi\bar{q}q}/M_{q}=1/f_{\pi}$
connecting the (constituent) quark mass parameter $M_q$ and the $\pi$-quark P-coupling strength $g_{\pi\bar{q}q}$ with $f_{\pi}$.

Such simple ``free" constituent quark loop (CQL) calculations are surprisingly successful. While in the present context the most important is their exact reproduction of the ``triangle" and ``box" anomalous amplitudes in (\ref{boxA}) and (\ref{ATAtheorem}), let us also recall that just the PVV quark triangle amplitude leads to over 15 decay amplitudes in agreement with data to within 3\% and not involving free parameters \cite{Delbourgo:1999qw,Delbourgo:1993dk,Kekez:2005kx}.
Since ``free" quarks here mean that there are no interactions between the effective constituent quarks in the loop, while they do couple to external fields, presently the photons $A_\mu$ and the pions $\pi_{a}$, the simplest CPT, Lorentz and $\mathrm{SU(2)}$ 
invariant 
effective Lagrangian encoding this is\footnote{The metric is given by $\eta_{11}=1$, $\eta_{22}=1$, $\eta_{33}=1$, $\eta_{00}=-1$}
\begin{equation}\label{1Leff}
\mathscr{L}_{\mathrm{eff}} = -\bar{q}(\slashchar{\partial} - ie\mathcal{Q}\,\slashchar{A} + M_{q} + 2ig_{\pi\bar{q}q}\gamma_{5}\pi+\ldots)q 
\, ,
\end{equation}
where $\pi=\pi_{a}t_{a}$, $t_{a}=\tau_{a}/2$ and $\tau_{a}$ Pauli matrices, 
whereas $\mathcal{Q}=\mbox{\rm diag}(\frac{2}{3},-\frac{1}{3})$ is the charge matrix of the quark iso-doublet $q=(u,d)^{T}$. The extension to $\mathrm{SU(3)}$ is obvious. The resulting CQL model calculation would be the same as, e.g., the lowest (one-loop) order calculation \cite{Bell:1969ts} in the quark--level linear $\sigma$-model \cite{Hakioglu:1990kg,Hakioglu:1991pn}.
Hence the ellipsis in the Lagrangian (\ref{1Leff}) -- to remind us that Eq. (\ref{1Leff}) also represents the lowest order terms pertinent for calculating photon-pion processes, from the $\sigma$-model Lagrangian and from all chiral quark model Lagrangians (e.g., see \cite{Andrianov:1998kj}) containing the mass term with the quark-meson coupling of the form
\begin{equation}
- M_q \, {\overline q}(UP_L+U^\dagger P_R)q \, ,
\label{massTerm}
\end{equation}
where $P_{L,R} \equiv (1\pm \gamma_5)/2$.
Namely, expanding
\begin{equation}
U^{(\dagger)} \equiv \exp[(-)i\pi/2f_\pi]
\end{equation}
to the lowest order in  
pion fields and invoking the GT relation, returns 
(\ref{1Leff}).

In contrast to this simple CQL model, a more sophisticated approach to 
quark-hadron physics is provided by the Dyson-Schwinger (DS) approach 
\cite{Alkofer:2000wg,Maris:2003vk,Fischer:2006ub}, 
which has clear connections with the underlying theory -- QCD. Namely, this approach clearly shows how the light pseudoscalar mesons simultaneously appear both as quark-antiquark ($q\bar q$) bound states and as Goldstone bosons of the dynamical chiral symmetry breaking (D$\chi$SB) of nonperturbative QCD, 
a unique feature among the bound--state approaches to mesons. Through D$\chi$SB in DS equation for quark propagators, dressed, momentum--dependent quark masses $M_q(p^2)$ are generated. They are in agreement with perturbative QCD for high momenta. 
However, thanks to D$\chi$SB, at low momenta they are of similar order of 
magnitude (and even tending to be somewhat higher) as typical constituent model mass parameters $M_q$. 
That is, $M_q\sim\frac{1}{3}$ of the nucleon mass $\sim\frac{1}{2}$ of the $\rho$-meson 
mass $m_\rho$, or higher if the mass defect due to the binding of quarks is taken into account.
This is true even in the chiral limit, i.e., for vanishing masses of fundamental quarks, 
which underscores the nonperturbative character of D$\chi$SB. The DS approach thus provides a partial justification of this simple CQL model, and adds to the understanding of its aforementioned phenomenological success \cite{Kekez:1998xr}.  
Namely, although the CQL model obviously suffers even from a lack of some qualitatively essential features, notably confinement, the assumption is that below spurious $\bar{q}q$ thresholds, a more important role is played by D$\chi$SB (which generates large constituent quark masses, i.e., 
$M_q \sim 300$--$500\mbox{ }\mathrm{MeV}$ for $u$ and $d$, 
the higher estimate being suggested by the DS approach; e.g., see Refs.
\cite{Alkofer:2000wg,Maris:2003vk,Fischer:2006ub} and references therein).

The DS approach uses the solutions of Bethe-Salpeter equations for the pseudoscalar meson bound--state vertices instead of the point pseudoscalar couplings $g_{\pi\bar{q}q}\gamma_{5}\tau_{a}$ of the CQL model (\ref{1Leff}). Then, both the 
anomalous amplitude $A_{\pi}^{2\gamma}$ and the connection (\ref{ATAtheorem}) with the box anomaly amplitude $A_{\gamma}^{3\pi}$ are again (in the chiral and soft limit) reproduced exactly and analytically \cite{Alkofer:1995jx,Bistrovic:1999dy} and independently of details of dynamics, which is again unique among the bound--state approaches. 

The extension of these amplitudes from the chiral and soft--point limits to general form factor kinematics have often been studied; e.g., Ref. \cite{Bistrovic:1999yy} used CQL (\ref{1Leff}) to study in this way the presently pertinent ``box" amplitude.
Present paper aims at continuing the study of Ref. \cite{Bistrovic:1999yy} by examining the possibility of including also the vector mesons, primarily in the description of the ``box"-anomalous transitions in a mixed quark-meson theory. As will become apparent below, this is a nontrivial and interesting theoretical issue in its own right, 
but there is also obvious 
phenomenological relevance in this context. 
For example, in the decays $\eta, \eta' \to 2\pi\gamma$, the vector mesons 
turn out to be essential for reproducing experimental results in very different approaches such as \cite{Venugopal:1998fq} and \cite{Delbourgo:1999qw}. Also, since in the process 
$\gamma^* \to \pi^+\pi^0\pi-$ (further, $\gamma^* \to 3\pi$ for short) 
one can depart strongly from low momenta, one may expect that the 
vector mesons will be important also here (e.g., see Ref. \cite{Holstein:1995qj}). 
On the other hand, if a treatment of a process is phenomenologically successful thanks to the inclusion of vector mesons (as in the cases $\eta, \eta' \to 2\pi\gamma$), an important question 
is whether anomalous processes are still described correctly in the low-energy 
(i.e., chiral and soft--point) limit. This problem has a somewhat lengthy history, of which only the presently necessary part will be reviewed in the beginning of the next section, which combines CQL
with vector mesons and finds the resulting $\pi^0 \to 2\gamma$ and 
$\gamma^* \to 3\pi$ 
amplitudes, revealing superfluous contributions of the resonant graphs
to $\gamma^* \to 3\pi$ in this limit.
In Sec. \ref{resolution}, we present a resolution of this problem.  In 
Sec. \ref{results}, we complete the calculation and discuss the results.
We summarize in Sec. \ref{summary}.

\section{Including Vector Mesons}
\label{vector}

\subsection{Short history of vector mesons in $\gamma^{*}\to3\pi$}

The Vector Meson Dominance (VMD) is certainly a reasonable approach to try 
because of its numerous empirical successes regarding electromagnetic interactions of hadrons (e.g., see \cite{O'Connell:1995wf} for a review and references), although its basis in 
QCD has not been fully clarified yet.
In a purely mesonic theory, Rudaz \cite{Rudaz:1974wt} assumed VMD where interaction with photons takes place only through $\rho^0$ and $\omega$ mesons. For example, $\pi^0\to 2\gamma$ would occur through $\pi^0\to\rho^0\omega$ and $\rho^0\to\gamma$, $\omega\to\gamma$ \cite{Ametller:1983ec}. 
Assuming the appropriate relationships between the pertinent coupling constants, 
he successfully reproduced the anomalous amplitude for $\pi^0\to 2\gamma$. Nevertheless, with the standard 
Kawarabayashi--Suzuki--Fayyazuddin--Riazuddin (KSFR)
relation~\cite{Kawarabayashi:1966kd,Riazuddin:1966sw} between $f_\pi$, $\rho$-meson mass $m_\rho$ and $\rho\pi\pi$ coupling $g_{\rho\pi\pi}$,
\begin{equation}
\frac{g_{\rho\pi\pi}^{2}}{m_{\rho}^{2}} \, = \, \frac{1}{2f_{\pi}^{2}},
\label{KSFR}
\end{equation}
this VMD approach (``pure" VMD in the following) would then give the amplitude $A_\gamma^{3\pi}$ too large by the factor of $\frac{3}{2}$, violating the axial anomaly relation 
(\ref{ATAtheorem}). To avoid this, he advocated \cite{Rudaz:1974wt} (in agreement with
Zinn-Justin and collaborators \cite{Basdevant:1969rw}) the KSFR relation revised 
by the factor $\frac{2}{3}$ \cite{Rudaz:1974wt}.
However, the experimental values of $g_{\rho\pi\pi}$, $m_{\rho}^{2}$ and $f_{\pi}$ strongly support (within few \%) the original one 
(\ref{KSFR}), which Rudaz finally adopted, also introducing \cite{Rudaz:1984bz} the contact term for the direct $\omega\to 3\pi$ transition. Namely, for a favorable choice 
of its coupling strength, this term could contribute $-\frac{1}{2}$ of the correct amplitude $A_\gamma^{3\pi}$, finally enabling the VMD approach to reproduce \cite{Rudaz:1984bz} the axial anomaly predictions (\ref{boxA}) and (\ref{ATAtheorem}). We denote this by ``modified" VMD. 
Subsequently, Cohen showed  \cite{Cohen:1989es} that the pertinent Ward-Takahashi 
identities (WI), first derived by Aviv and Zee \cite{Aviv:1971hq}, support the 
existence of such an extra contact term.

\subsection{CQL--VMD models}

We want to examine whether anomalous processes like $\pi^{0}\to 2\gamma$ and $\gamma^{*}\to 3\pi$ can be properly described by a mixed model of constituent quarks and mesons, which, unlike the Lagrangian (\ref{1Leff}), would include not only pseudoscalar, but also vector mesons. Thus, the quark--meson--interaction part of the 
Lagrangian (\ref{1Leff}) gets enlarged to 
\begin{equation}\label{inc1}
\mathscr{L}_{\mathrm{int}} = -2ig_{\pi\bar{q}q}\bar{q}\gamma_{5}\pi q + ig_{\rho\bar{q}q}\bar{q}\gamma^{\mu}\rho_{\mu}q,
\end{equation}
where\footnote{We assumed the ideal $\omega$--$\phi$ mixing, as well as an $\mathrm{U(2)}$ symmetry for the interactions of quarks and vector mesons.} $\rho^{\mu}=\rho^{\mu}_{a}t_{a}+\omega^{\mu} t_{0}$, $t_{0}=\tau_{0}/2$, $\tau_{0}=\mathrm{diag}(1,1)$. This quark-meson interaction is, for instance, used in the quark-loop approach of Refs. \cite{Delbourgo:1999qw,Delbourgo:1993dk,Kekez:2005kx} -- a prominent example, since, as pointed out earlier, it describes many processes without involving free parameters.

\begin{figure}[htb!]
\includegraphics[scale=0.8]{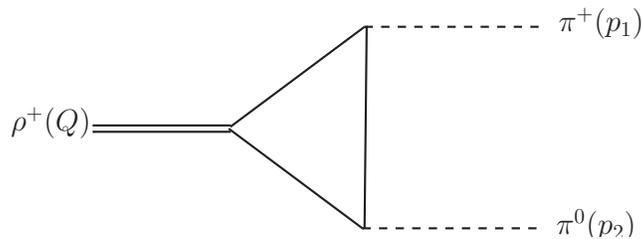}
\caption{The form factor $G_{\rho\pi\pi}$ seen as a quark triangle diagram in the 
CQL model.}
\label{rho_triangle}
\end{figure}

With this theory at hand, one could compute the $G_{\rho\pi\pi}$ form factor in the quark-loop model, where the $\rho\to\pi\pi$ amplitude is represented by a quark triangle 
(Fig. \ref{rho_triangle})
, and its crossed mate, with VPP couplings.  In terms of an effective Lagrangian, the $\rho\pi\pi$ interaction would then be given by
\begin{equation}\label{inc2}
\mathscr{L}_{\rho\pi\pi} = 2 i G_{\rho\pi\pi}\mathrm{Tr}([\pi,(\partial_{\mu}\pi)]\rho^{\mu}),
\end{equation}
which implies that besides the ``box" VPPP graphs in Fig. \ref{figbox}, 
the process $\gamma^* \to 3\pi$ also receives contributions from the 
$\rho$-resonant triangle graphs such as the one in Fig. \ref{r_triangle}.

For $G_{\rho\pi\pi}$ we use the chiral and soft--point limit result by Hakioglu and Scadron \cite{Hakioglu:1991pn}: $\lim_{m_{\pi\to 0}} G_{\rho\pi\pi}(Q=0) \equiv g_{\rho\pi\pi} = g_{\rho\bar{q}q} = \mathrm{const.}$, consistent with the hypothesis of the VMD universality, which assumes the equality of all $\rho$ couplings, namely the couplings to fermions (presently -- quarks $q$), to pions $\pi$, and to photons $\gamma$, i.e., 
\begin{equation}\label{universality}
g_{\rho\bar{q}{q}}=g_{\rho\pi\pi}=g_{\rho\gamma} = \mathrm{const}.=g_\rho \, .
\end{equation}

\begin{figure}[b!]
\includegraphics[scale=0.8]{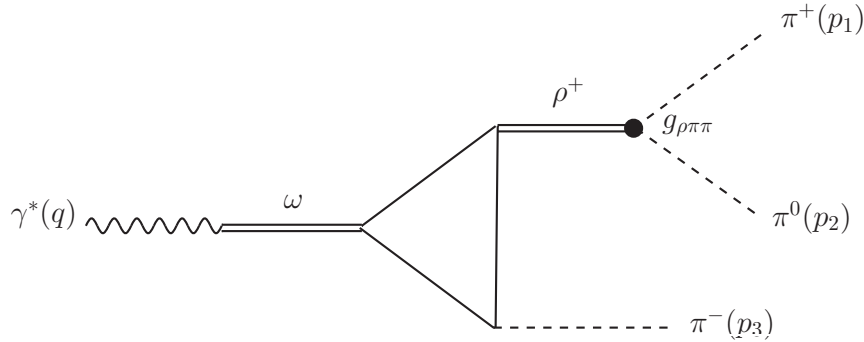}
\caption{One of the ``resonant" triangle diagrams for the process $\gamma(q) \to \pi^+(p_1) \pi^0(p_2) \pi^-(p_3)$, where the two pions with momenta $p_1$ and $p_2$ are obtained by the decay of the intermediate $\rho$ meson transferring the momentum $Q = p_1 + p_2$. Two more analogous graphs are obtained by replacing $\rho^+ \to \pi^+ \pi^0$ and $\pi^-$ by, respectively, $\rho^0 \to \pi^+ \pi^-$ and $\pi^0$, or by $\rho^- \to \pi^-\pi^0$ and $\pi^+$. Also, each of the three graphs has its crossed graph. } 
\label{r_triangle}
\end{figure}

Electromagnetism is already present in the starting Lagrangian (\ref{1Leff}) through the direct quark--photon coupling $ie\bar{q}\mathcal{Q}\slashchar{A}q$, but to incorporate VMD in our theory, 
the coupling of the photon to neutral vector mesons must be added.
One way to describe the interactions between photons and hadronic matter in the spirit of VMD is
\begin{equation}\label{inc2a}
\mathscr{L}_{\mathrm{VMD1}}=ie\bar{q}\mathcal{Q}\gamma^{\mu}A_{\mu}q - \frac{e}{2}F^{\mu\nu}\Bigl(\frac{1}{g_{\rho\gamma}}\rho^{0}_{\mu\nu}+
\frac{1}{g_{\omega\gamma}}\,\omega_{\mu\nu}\Bigr) \, .
\end{equation}
This version of VMD is often called VMD1.   Here 
$F_{\mu\nu}=\partial_{\mu}A_{\nu}-\partial_{\nu}A_{\mu}$,
$\omega_{\mu\nu}=\partial_{\mu}\omega_{\nu}-\partial_{\nu}\omega_{\mu}$, 
and also
$\rho^{0}_{\mu\nu}=\partial_{\mu}\rho^{0}_{\nu}-\partial_{\nu}\rho^{0}_{\mu}$ 
\cite{O'Connell:1995wf}. 
Ideal $\omega$--$\phi$ mixing gives $g_{\omega\gamma}=3g_{\rho\gamma}$. VMD1 can be transformed into the most popular representation of VMD where there is no direct quark--photon coupling. In the limit of universality (\ref{universality}), 
this standard (``Sakurai's") representation of VMD, denoted VMD2, is 
equivalent \cite{O'Connell:1995wf} to VMD1. The presently pertinent part of the VMD2 Lagrangian reads\footnote{Where we omit the kinetic term and the fictitious ``photon mass term" which arises (e.g., see \cite{O'Connell:1995wf}) from the gauge invariance of the theory, which is manifest in VMD1 (\ref{inc2a}).}
\begin{equation}\label{inc3}
\mathscr{L}_{\mathrm{VMD2}}=\frac{em_{\rho}^{2}}{g_{\rho\gamma}} \, A^{\mu} \, \Bigl(\rho^{0}_{\mu}+\frac{1}{3}\, \omega_{\mu}\Bigr).
\end{equation}
Our starting CQL Lagrangian (\ref{1Leff}) is thus finally augmented to include vector mesons (further denoted as the CQL--VMD model):
\begin{equation}\label{2Leff}
\mathscr{L}_{\rm eff}=\mathscr{L}_{\mathrm{int}}+\mathscr{L}_{\mathrm{VMD2}}+\mathscr{L}_{\rho\pi\pi}-\bar{q}(\slashchar{\partial} + M_{q} + ... )q 
\, .
\end{equation}
The ellipsis in Eq. (\ref{2Leff}) again serve to remind one that the terms not pertinent for photon--pion processes are not shown. (E.g., if the starting Lagrangian (\ref{1Leff}) was the sigma-model one, we would have terms containing scalar, $\sigma$-mesons also in Eq. (\ref{2Leff}), but they do not contribute to the presently interesting $\gamma^{*}\to 3\pi$ and $\pi^{0}\to 2\gamma$. That is, the contribution of the graphs analogous to the resonant triangles, but with intermediate $\sigma\to\pi^{+}\pi^{-}$ instead of $\rho^{0}\to\pi^{+}\pi^{-}$, vanishes due to parity conservation.) These terms are, in general, different for different theories, but since they are presently irrelevant, our conclusions will be the same for many different models, from the $\sigma$-model
to the chiral quark models \cite{Andrianov:1998kj}, after the $\rho$ and $\omega$ mesons are introduced to them. In that sense, CQL--VMD denotes not just one, but the whole class of models. 

In contrast to $\mathscr{L}_{\mathrm{VMD1}}$ (\ref{inc2}), there is no direct quark--photon coupling in the VMD2 picture (\ref{2Leff}). Thus, in VMD2, the six box graphs exemplified by Fig.~\ref{figbox} 
do not contribute to $\gamma^{*}\to 3\pi$ in the precise form depicted in Fig.~\ref{figbox}. Instead, they are modified so that the photon first couples to the intermediate $\omega$-meson which in turn couples to quarks.
That is, the quark-photon coupling in Fig. \ref{figbox} is replaced by the photon coupling $g_{\omega\gamma}$ to $\omega$ propagating to 
its $q\bar q$ vertex.  There is no need to spend space on re--drawing Fig.~\ref{figbox} to depict this insertion of $\omega$ since it modifies the photon coupling in the same way in all graphs, and is illustrated in Fig.~\ref{r_triangle}. (Note that in VMD1, there are both the graphs with the direct quark--photon coupling and their partners with the $\omega$-insertion, but since the momentum dependence is different than in VMD2, the sum of these graphs in VMD1 yields the same results as VMD2.)

The box and resonant triangle graphs in which the photon is coupled to $\rho^0$ (instead of to $\omega$), all vanish, as must be due to  G-parity conservation.

\subsection{$\pi^{0}\to 2\gamma$ through CQL--VMD models}

Since we choose to work with VMD2, the outgoing photons in this process are created only through the mediation with the $\rho^{0}$ and $\omega$. In our model, these, in turn, come from a ``triangle" PVV quark loop. If the photons are on-shell there is a complete cancellation of the $\omega$ and $\rho$ propagators with $m_{\rho}^{2}$ in the VMD coupling. In the same manner, and by using universality, the $g_{\rho\bar{q}q}$ and $g_{\omega\bar{q}q}$ couplings get canceled with the $g_{\rho\gamma}$ and $g_{\omega\gamma}$, respectively. Then, our Lagrangian (\ref{2Leff}) leads to the same $\pi^{0}(p)\to\gamma(k_{1})\gamma(k_{2})$ amplitude as the standard quark--triangle--loop calculation (e.g., see \cite{Ametller:1983ec}), namely
\begin{equation}\label{inc5}
\mathscr{M}_{\pi}^{2\gamma} \, = \, k_{1\mu} \, \epsilon_{\nu}^{\ast}(\mathbf{k}_{1},\sigma_{1})k_{2\rho} \, \epsilon_{\lambda}^{\ast}(\mathbf{k}_{2},\sigma_{2})
 \, \varepsilon^{\mu\nu\rho\lambda} \, T_{\pi}(k_{1},k_{2}),
\end{equation}
where $k_i$ and $\sigma_i$ are the momentum and polarization of $\gamma(k_i)$, and 
\begin{equation}
T_{\pi}(k_{1},k_{2}) \, = \, \frac{e^{2}N_{c}}{12\pi^{2}} \, \frac{g_{\pi\bar{q}q}}{M_{q}}\, \widetilde{C}_{0}(k_{1},k_{2})\, .
\end{equation}
Here $\widetilde{C}_{0}=(2!M^{2}/i\pi^{2})C_{0}$, where $C_{0}$ is the standard 't Hooft--Veltman \cite{'tHooft:1978xw} scalar three--point function. The limit $m_{\pi}\ll M_{q}$, together with the GT relation, reproduces the analytical result
\begin{equation}\label{inc6}
\lim_{m_{\pi}\to 0}T_{\pi}=\frac{e^{2}N_{c}}{12\pi^{2}}\, \frac{1}{f_{\pi}} \, = \, A_{\pi}^{2\gamma}
\end{equation}
for the anomalous chiral $\pi^0$ decay into two real photons, $k_1^2=0,k_2^2=0$.

\subsection{The problem with $\gamma^{*}\to 3\pi$ in CQL--VMD models}

Previous short calculation served as a consistency check aimed at reproducing the correct low energy limit even when vector mesons are introduced. Here we perform a similar calculation for 
$\gamma^{*}(q)\to\pi^{+}(p_{1})\pi^{0}(p_{2})\pi^{-}(p_{3})$, and obtain the amplitude
\begin{equation}\label{inc7}
\mathscr{M}^{3\pi}_{\gamma}\, =\, \epsilon_{\mu}(\mathbf{q},\sigma)\, p_{1\nu}\, p_{2\rho}\, p_{3\lambda}\, \varepsilon^{\mu\nu\rho\lambda}
\, F_{\gamma}^{3\pi}(p_{1},p_{2},p_{3}),
\end{equation} 
In this notation, the contribution from the resonant triangles reads
\begin{multline}\label{inc9}
F_{\triangle}^{\mathrm{res}}(p_{1},p_{2},p_{3})=
\frac{1}{2}\frac{eN_{c}}{6\pi^{2}}\frac{g_{\pi\bar{q}q}}{M_q}\frac{g_{\rho\bar{q}q}^{2}}{m_{\rho}^{2}}
\frac{m_{\rho}^{2}}{m_{\rho}^{2}+q^{2}}\\
\times\Bigl[\frac{m_{\rho}^{2}}{m_{\rho}^{2}-s}\widetilde{C}_{0}(p_{1},p_{2})+
\frac{m_{\rho}^{2}}{m_{\rho}^{2}-t}\widetilde{C}_{0}(p_{2},p_{3})+\frac{m_{\rho}^{2}}{m_{\rho}^{2}-u}\widetilde{C}_{0}(p_{1},p_{3})\Bigr].
\end{multline}
where the Mandelstam variables are defined{\footnote{Note that this definition 
is different from the convention in the Serpukhov paper \cite{Antipov:1986tp}
and, e.g., Ref. \cite{Ametller:2001yk}. Our notation is closest to that of
the proposal of the CEBAF experiment \cite{Miskimen94}, where 
the outgoing pion pair is $\pi^+\pi^0$. Thus the choice
$s=-(p_{1}+p_{2})^{2}=-(p_{\pi^+}+p_{\pi^0})^{2}$, while the squared invariant 
mass of the pion pair outgoing in the Serpukhov experiment, $\pi^-\pi^0$, is 
$t=-(p_{2}+p_{3})^{2}=-(p_{\pi^0}+p_{\pi^-})^{2}$ (see Fig. \ref{figbox}). 
}}
 by $s=-(p_{1}+p_{2})^{2}$, $t=-(p_{2}+p_{3})^{2}$, $u=-(p_{1}+p_{3})^{2}$. 
The form factor from the box graphs of Fig. \ref{figbox}, is
\begin{multline}\label{inc10}
F_{\Diamond}(p_{1},p_{2},p_{3})=\frac{1}{3}\frac{eN_{c}}{6\pi^{2}}\frac{g_{\pi\bar{q}q}}{M_{q}}\frac{g_{\rho\bar{q}q}^{2}}{m_{\rho}^{2}}
\frac{m_{\rho}^{2}}{m_{\rho}^{2}+q^{2}}\\
\times\Bigl[\widetilde{D}_{0}(p_{1},p_{2},p_{3})+\widetilde{D}_{0}(p_{1},p_{3},p_{2})+\widetilde{D}_{0}(p_{2},p_{1},p_{3})\Bigr]
\end{multline}
where $\widetilde{D}_{0}=(3!M^{4}/i\pi^{2})D_{0}$ and $D_{0}$ is the 't Hooft--Veltman scalar four--point function \cite{'tHooft:1978xw}. 
The total amplitude would be
$F_{\gamma}^{3\pi}=F_{\triangle}^{\mathrm{res}}+F_{\Diamond}$.

In the soft--point and chiral limit (i.e. $p_{1},p_{2},p_{3}\to 0$), where $\widetilde{C}_{0},\widetilde{D}_{0}\to 1$, and with the usage of GT and KSFR relations, 
\begin{equation}\label{inc11}
F_{\triangle}^{\mathrm{res}}\to\frac{3}{2}\, \frac{eN_{c}}{12\pi^{2}}\frac{1}{f_{\pi}^{3}}
= \, \frac{3}{2} \, A_{\gamma}^{3\pi} ,
\qquad 
F_{\Diamond}\to\frac{eN_{c}}{12\pi^{2}}\frac{1}{f_{\pi}^{3}} \, = \, A_{\gamma}^{3\pi},
\end{equation}
which would mean a total of
$F_{\gamma}^{3\pi}(0,0,0)\to\, 
\frac{5}{2}\, \frac{eN_{c}}{12\pi^{2}}\frac{1}{f_{\pi}^{3}}
= \, \frac{5}{2} \, A_{\gamma}^{3\pi}$,
which is by the factor 5/2 bigger than the correct anomalous amplitude.
Notice that if we had included only the resonant triangles we would have a result that is off by $3/2$; these are the very same $3/2$ that we mentioned 
after Eq. (\ref{KSFR}) as the reason for Rudaz adding the $\omega-3\pi$ contact term into the previously ``pure" VMD description.
It appears that the above Lagrangian (\ref{inc1}), (\ref{inc3}) leads to an inconsistency; if we want to calculate anomalous processes with more than one pseudoscalar, it is not legitimate to just add vector mesons \`{a} la Sakurai
to a mixed meson--CQL. It is easy to check by explicit calculation that the same problem persists (as expected) if VMD1 approach is used instead.

\section{A resolution through Weinberg--Tomozawa interaction}
\label{resolution}

In the present context of strong interactions and hadrons, which, after all, have substructure, one may think 
of introducing a form factor $\mathscr{F}^V(Q^2)$ for the transition from the vector quark vertex to $2\pi$ 
 instead of  the corresponding part of the resonant triangle graphs (Fig. \ref{r_triangle}), which makes the latter troublesome. Such a form factor would be constrained by the known anomalous behavior of the {\it total} form factor $F_\gamma^{3\pi}(p_1,p_2,p_3)$ in the soft point. However, introducing it by hand and {\it ad hoc}, without an insight into the underlying dynamics, would not be a satisfactory way of removing the superfluous contributions. To understand which modifications of the CQL--VMD approach to make, we seek guidance from a more fundamental, substructure level. 

\begin{figure}[htb!]
\includegraphics[scale=0.8]{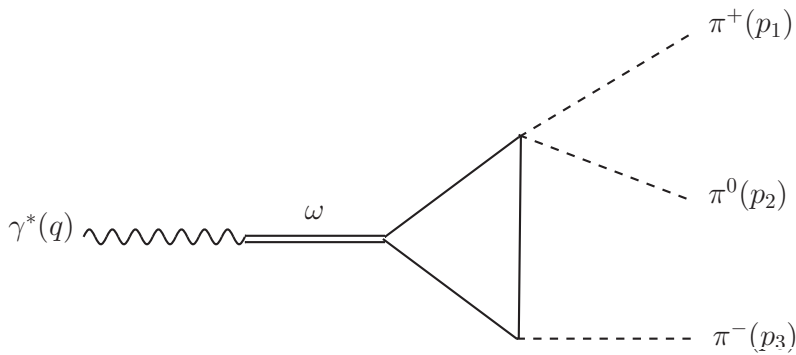}
\caption{One of the graphs for the $\gamma^{*}\to 3\pi$ non-resonant triangle contribution depicting a $\pi^{+}\pi^{0}$ pair coming out from the $\bar{q}q\pi\pi$ Weinberg-Tomozawa (WT) interaction 
that is needed for the appropriate subtraction. The two analogous graphs are obtained by exchanging the pion pair by $\pi^{+}\pi^{-}$ and $\pi^{0}\pi^{-}$. There are two of each of these graphs.}
\label{nr_triangle}
\end{figure}

\subsection{Insight from the Dyson-Schwinger approach}

In the Introduction we already mentioned the DS approach to QCD.
Although there is no full derivation of CQL or VMD from QCD as the underlying theory, much of their 
features can be reproduced and understood in the process of describing pseudoscalar and vector mesons and their interactions in the DS approach, which is free from the above problem of the superfluous contributions $\frac{3}{2} \, A_{\gamma}^{3\pi}$ (\ref{inc11})
of the resonant graphs to the anomalous amplitude. 

The DS approach employs dynamically dressed quark propagators 
$S(k) = [{\rm i} \rlap{$k$}/ \, A(k^2) + B(k^2)]^{-1}$ obtained by solving 
the ``gap" DS equation, so that the momentum-dependent mass function 
$M_q(k^2)\equiv B(k^2)/A(k^2)$ takes place of the simple constant constituent 
mass $M_q$. Also, the solutions of the Bethe-Salpeter (BS) equations for the 
pseudoscalar meson bound--state vertices replace the point pseudoscalar 
couplings $g_{\pi\bar{q}q}\gamma_{5}$ of the CQL model (\ref{1Leff}), and a 
vector WI-preserving dressed quark-photon vertex $\Gamma_\mu(k,k')$ 
is in place of the bare quark-photon vertex $\gamma_\mu$.
(In the vast majority of phenomenological applications this is 
the Ball-Chiu {\it Ansatz} \cite{Alkofer:2000wg,Maris:2003vk}.)
This procedure defines the generalized impulse approximation (GIA). 
As we already mentioned, this reproduces the $A_{\pi}^{2\gamma}$ and 
$A_{\gamma}^{3\pi}$ low energy theorems. 

On the other hand, the momentum dependence, i.e., the growth of the form factor $F_\gamma^{3\pi}(p_{1},p_{2},p_{3})$ from its soft--point limit  $A_{\gamma}^{3\pi}$ turned out to be so slow (after all six permutations of the graph in Fig. \ref{figbox} were taken into account properly \cite{Bistrovic:1999dy}), that the Serpukhov data point could not possibly be explained in that approach. 
(Also the related processes $\eta, \eta' \to 2\pi\gamma$ could hardly be reconciled with such a weak momentum dependence \cite{Bistrovic:1999dy} of the  $\gamma^{*} \to 3\pi$ form factor.) This indicated that the DS approach should include the contributions from vector mesons -- or rather, in the DS context, their microscopic, $q\bar q$--substructure equivalent. To this end, Ref. \cite{Cotanch:2003xv} went beyond GIA in its treatment of the box graph, inserting and summing the infinite set of gluon ladder exchanges in the $s, t$ and $u$--channel of the box graph. The resulting momentum dependence of $F_\gamma^{3\pi}(p_{1},p_{2},p_{3})$ agrees well with the VMD behavior; the inclusion of these gluon diagrams beyond GIA successfully produced $\rho$--meson--like intermediate states in the two--pion channels. Thus, their contributions correspond to our resonant triangle graphs, but with the crucial difference that in the DS case these contributions vanish as one approaches the soft limit of vanishing momenta, so that the correct anomalous amplitude is obtained, unlike Eq. (\ref{inc11}).

This favorable behavior can be understood on the basis of Maris and Tandy's DS results on dressed vector vertices \cite{Maris:1999bh}. In the DS approach one does not have elementary meson fields, but one explicitly constructs physical, on-mass-shell mesons as $q\bar q$ bound states which are eigenstates of mass. Thus, the mesons are not well-defined away from their mass poles, nor are their couplings (such as $g_{\rho\pi\pi}, g_{\rho\gamma}$ and $g_{\rho\bar{q}q}$). Therefore, at the level of meson substructure, the issue of the $\rho$-meson-like intermediate states, including the resonant $\rho$ contributions to the two-pion channels, should be addressed within the dressed quark vector vertex $\Gamma_\mu(k,k')$, which couples not only to photons but also to the pion vector current $\mathrm{Tr}([\pi,(\partial_{\mu}\pi)]t_{a}\,)$.

Using the same DS dynamical model{\footnote{This model is the most widely used one 
in the phenomenological branch of DS studies -- see DS approach reviews 
such as Refs. \cite{Alkofer:2000wg,Maris:2003vk,Fischer:2006ub}} 
for the quark-gluon interactions as later Ref. \cite{Cotanch:2003xv}, Maris and Tandy \cite{Maris:1999bh} solved this inhomogeneous BS equation for the dressed quark vector vertex $\Gamma_\mu(k,k')$. In this way they essentially reconciled the VMD picture with the QCD picture of a photon coupled to quarks in a $q\bar q$ bound state; for the present paper their most important result was that the inhomogeneous BS equation generated both resonance and nonresonance contributions to the full (model) vector vertex $\Gamma_\mu(k,k')$, which, unlike the BC vertex, contains timelike vector-meson pole (at the model value $Q^2 = -m_V^2 = -m_\rho^2 = -m_\omega^2 \approx -0.55$ GeV$^2$) in the part of the vertex transverse to $Q_\mu$. While this part (the resonant part) of the vector vertex BS solution is significantly enhanced over the BC {\it Ansatz}, it also vanishes as $Q^2 \to 0$. This is explicitly shown in Eqs. (30) and (34) of Ref. \cite{Maris:1999bh}. 

\subsection{Subtraction of the Weinberg-Tomozawa interaction}

In terms of meson degrees of freedom, this means that the resonant contribution from the intermediate vector meson with $Q^2=0$ is absent. This is reasonable as it would correspond to a constant meson propagator $1/m_V^2$, and this in turn corresponds to a point interaction in the coordinate space. In the present case, it would correspond to a $\rho$-meson propagating {\it zero} distance from its $q\bar q$ vertex before turning into two pions, i.e., to a $q\bar q$ vertex producing two pions {\it immediately}, since here $Q^2$ is either $s$, $t$, or $u$, depending on the two-pion channel coupled to $\rho$. Note that this unphysical situation is quite different from the situation when such ``intermediate but non-propagating" $\rho$ turns into a $\gamma$: this just means that also VMD2 incorporates {\it implicitly} the situation when the photon couples to quarks immediately and directly (for example, as in the case of $\pi^{0}\to2\gamma$, where the factors of $m_\rho$ and couplings, except $e$, cancel). Contrary to that, two pions can come from a vector $q\bar q$ vertex only via a truly propagating intermediate $\rho$ meson (with transferred momentum $Q^2\neq 0$), while there is no direct conversion of quarks and antiquarks into the two-pion vector current.

This is the reason why our previous calculation has led to the spurious soft--point contribution $\frac{3}{2}\, A_{\gamma}^{3\pi}$ (\ref{inc11}): in the resonant graphs, the part 
$$\bar{q}\gamma^{\mu}t_{a} q \to {\rm intermediate} \, \rho_{a}^{\mu} \to  \epsilon_{abc} \pi_{b}\partial^{\mu}\pi_{c}$$
yields simply{\footnote{ -- thanks to vanishing of the transverse part of $\rho$ propagators when contracted with Levi--Civitas from the traces of PVV triangles. The second equality is from the KSFR relation.}} the ``vector $q\bar q \to 2\pi$ form factor"
\begin{equation}
\mathscr{F}_{\rho}^V(Q^2) \propto g_{\rho q\bar q} \, \frac{1}{Q^2 + m_\rho^2} \, g_{\rho\pi\pi} 
= \frac{m_\rho^2}{2 f_{\pi}^{2}} \, \frac{1}{Q^2 + m_\rho^2} \, ,
\label{wrongFF}
\end{equation}
which is, nevertheless, wrong as it stands because it contains the contribution of the intermediate $\rho$ with $Q^2 = 0$.  
The correct vector $q\bar q \to 2\pi$ form factor can be obtained by subtracting this contribution:
\begin{equation}
\mathscr{F}^V(Q^2) = \mathscr{F}_{\rho}^{V}(Q^{2})- \mathscr{F}_{\rho}^{V}(0)
\propto \, g_{\rho q\bar q} \, \frac{1}{Q^2 + m_\rho^2} \, g_{\rho\pi\pi}
       -  g_{\rho q\bar q} \, \frac{1}{m_\rho^2} \, g_{\rho\pi\pi}
= \frac{1}{2 f_{\pi}^{2}} \, \frac{- Q^2}{Q^2 + m_\rho^2} \, ,
\label{corrFF}
\end{equation}
i.e., the resonant contributions depend on $Q^2$ essentially as in the DS substructure considerations such as \cite{Cotanch:2003xv}
and \cite{Maris:1999bh} (see esp. Eqs. (30) and (34)).

One may visualize the removal of the point $q\bar q \to 2\pi$ interaction (non-propagating, $Q^2=0$ $\rho$) leading to Eq. (\ref{corrFF})
as the subtraction of the point-interaction triangle graphs (such as Fig. \ref{nr_triangle}) from the corresponding resonant triangles (e.g., Fig. \ref{r_triangle}).  In terms of formulas, this subtraction corresponds to including the following $\pi\pi\bar{q}q$ point coupling into the effective $\rho\pi\pi$ part like this:
\begin{equation}
2 i g_{\rho\pi\pi}\mathrm{Tr}([\pi,(\partial_{\mu}\pi)]\rho^{\mu}) \to
2 i \mathrm{Tr}([\pi,(\partial_{\mu}\pi)](g_{\rho\pi\pi}\rho^{\mu}-ig_{\pi\pi\bar{q}q}J^{\mu}))
\label{additional}
\end{equation}
where $J^{\mu}=t_{a}J_{a}^{\mu}$, $J_{a}^{\mu}=\bar{q}\gamma^{\mu}t_{a}q$, and $g_{\pi\pi\bar{q}q}$ 
is fixed precisely in a way to respect the $\gamma^{*}\to 3\pi$ low energy theorem, i.e., 
in a way that it cancels the resonance part in the soft--point limit \textit{completely}, 
yielding $g_{\pi\pi\bar{q}q} = \frac{1}{2 f_{\pi}^{2}}$. Written in this form, it turns out to be 
nothing else but the quark-level Weinberg-Tomozawa (WT) interaction \cite{Weinberg:1966zz,Tomozawa:1966jm}.
However, it must be understood that Eq. (\ref{additional}) 
indicates only the subtraction of the resonant graphs and does not mean adding a new $\pi\pi\bar{q}q$ interaction term to the Lagrangian (\ref{2Leff}). If one would try this, the KSFR relation would be spoiled in the same way as when the analogous two-pion interaction with nucleons is added to the VMD-nucleon Lagrangian \cite{Cohen:1989es}.

This subtraction, that needs to be included only when dealing with PVV triangles coupled to vector mesons decaying to two pseudoscalars, 
completes the definition of our constituent quark model coupled to pseudoscalar and vector mesons.  With it, the combination of the resonant triangles (e.g., Fig. \ref{r_triangle}) and the ``subtraction triangles" with the added point $\pi\pi\bar{q}q$ interaction (e.g., Fig. \ref{nr_triangle}) yields the behavior in accord with the Abelian axial anomaly of QCD, as shown in detail in the beginning of the next section, where we complete the calculation of the $\gamma^* \to 3\pi$ form factor.

\section{Results and discussion}
\label{results}

\subsection{Completing the calculation}
\label{completingCalc}

When we include the non--resonant, WT triangles (Fig. \ref{nr_triangle}) in the amplitude, 
the complete triangle form--factor 
$F_{\triangle}=F_{\triangle}^{\mathrm{res}}+F_{\triangle}^{\mathrm{WT}}$ 
becomes
\begin{multline}\label{cal1}
F_{\triangle}(p_{1},p_{2},p_{3}) = \, \frac{1}{2} \,\, 
\frac{eN_{c}}{6\pi^{2}}\frac{g_{\pi\bar{q}q}}{M_q}\frac{g_{\rho\bar{q}q}^{2}}{m_{\rho}^{2}} \, \, 
\frac{m_{\rho}^{2}}{m_{\rho}^{2}+q^{2}}\\
\times\Bigl[\frac{s}{m_{\rho}^{2}-s} \, \widetilde{C}_{0}(p_{1},p_{2}) +
\frac{t}{m_{\rho}^{2}-t} \, \widetilde{C}_{0}(p_{2},p_{3}) + 
\frac{u}{m_{\rho}^{2}-u} \, \widetilde{C}_{0}(p_{1},p_{3})\Bigr].
\end{multline}

With the usage of GT and KSFR relations, the constant prefactor becomes
\begin{equation*}
\frac{eN_{c}}{6\pi^{2}}\frac{g_{\pi\bar{q}q}}{M_q}\frac{g_{\rho}^{2}}{m_{\rho}^{2}}
=\frac{eN_{c}}{12\pi^{2}f_\pi^3} = A_{\gamma}^{3\pi} .
\end{equation*}
\vskip 3mm

The total amplitude is $F_{\gamma}^{3\pi}=F_{\triangle}+F_{\Diamond}$. In the soft--point limit (i.e. $p_{1},p_{2},p_{3}\to 0$), 
where $\widetilde{C}_{0},\widetilde{D}_{0}\to 1$, $F_{\triangle} = F_{\triangle}^{\mathrm{res}} + F_{\triangle}^{\mathrm{WT}}\to 0$
and the total is in this limit given by the pseudoscalar box contribution
\begin{equation*}
F_{\gamma}^{3\pi}(p_{1},p_{2},p_{3}\to 0) \to F_{\Diamond}(p_{1},p_{2},p_{3}\to 0) \to\frac{eN_{c}}{12\pi^{2}}\frac{1}{f_{\pi}^{3}} 
= A_{\gamma}^{3\pi}.
\end{equation*}

It is interesting to note that for real photons ($q^{2}=0$), and by ``squeezing" the quark triangles and boxes to points, the total form factor $F_{\gamma}^{3\pi}=F_{\Diamond}+F_{\triangle}$ reads
\begin{equation}\label{cal2}
F_{\gamma}^{3\pi}\to A_{\gamma}^{3\pi}\Bigl[1+\frac{1}{2}\Bigl(\frac{s}{m_{\rho}^{2}-s}+\frac{t}{m_{\rho}^{2}-t}+\frac{u}{m_{\rho}^{2}-u}\Bigr)\Bigr].
\end{equation}
where we used the GT and KSRF relation. This is precisely the Terent'ev phenomenological form \cite{Terent'ev:1971kt,Terentev:1974vr} for $\delta=0$ and $C_{\rho}=1/2$ (in other words, a ``modified" VMD result \cite{Rudaz:1984bz,Holstein:1995qj}).

\begin{figure}[b!]
\includegraphics[scale=0.9]{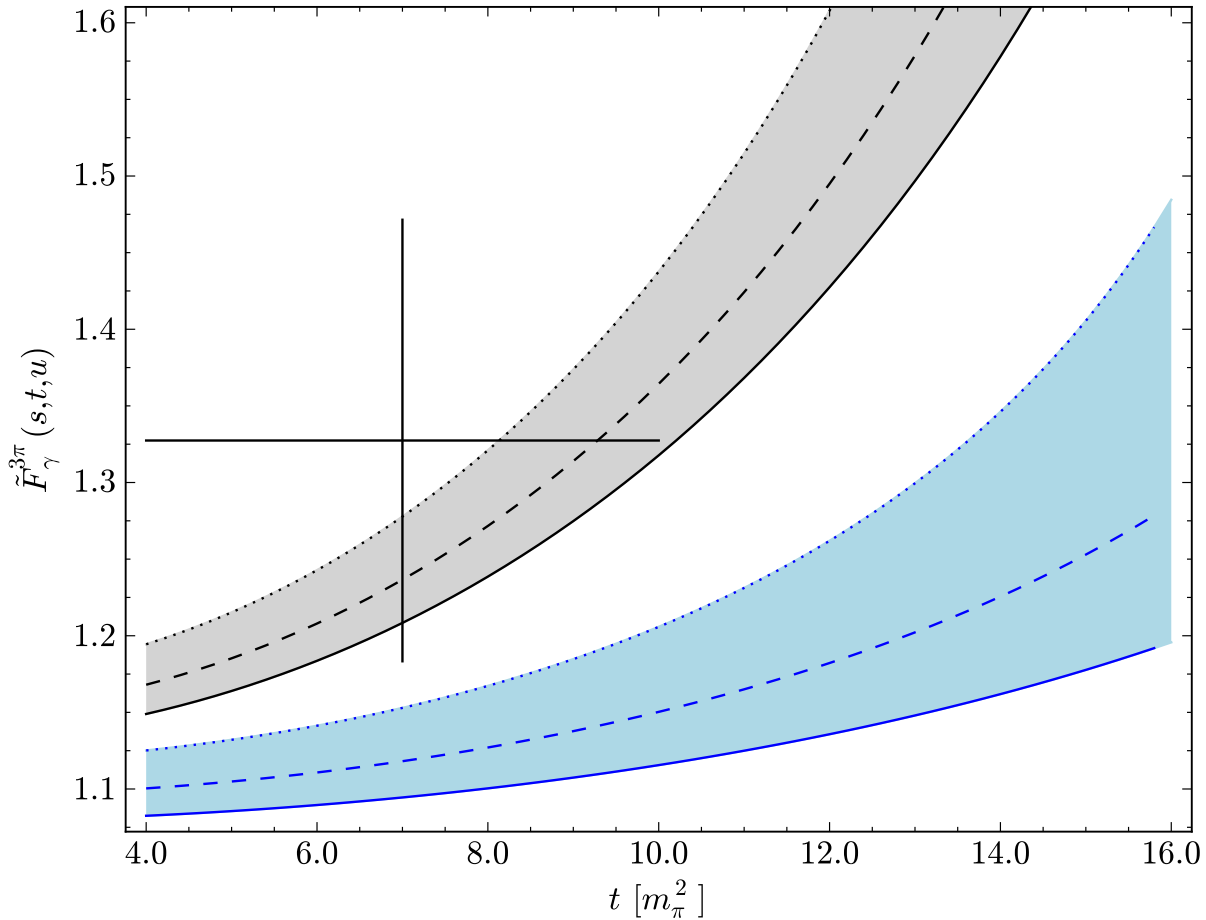}
\caption{(color online) 
For various values of the 
mass parameter $M_q$, 
our numerically calculated form factors $\tilde{F}_{\gamma}^{3\pi}(s,t,u)$ 
are shown as functions of $t=-(p_2+p_3)^2$ $=-(p_{\pi^0}+p_{\pi^-})^2$,
i.e., of the invariant mass of the outgoing pion pair $\pi^0\pi^-$ in the Serpukhov 
and COMPASS experiments. There, all pions are on shell, and for definiteness
we fix $u=m_{\pi}^{2}$. 
The curves belonging to the lower stripe (blue online) are only the quark box contribution
(i.e., the lower stripe represent predictions of the simple CQL approach for various $M_q$). 
The upper stripe (with black curves) represents the corresponding predictions from the
presently pertinent CQL--VMD approach;
that is, the black curves represent predictions which include the $\rho$-resonant triangle 
loops with the subtracted Weinberg-Tomozawa interaction. In the both stripes, 
the dotted curves correspond to $M_{q}=300$ MeV, the dashed ones to $M_{q}=330$ MeV 
and the solid ones to $M_{q}=360$ MeV. 
The exhibited data point \cite{Antipov:1986tp}  
really corresponds to the average value of the form factor over the 
momentum range covered by the experiment 
(between the two-pion threshold and 10 $m_\pi^2$),
which measured the total cross-section -- see 
the discussion in Subsec. \ref{ComparWexperiment}.}
\label{prim1}
\end{figure}

\begin{figure}[b!]
\includegraphics[scale=0.9]{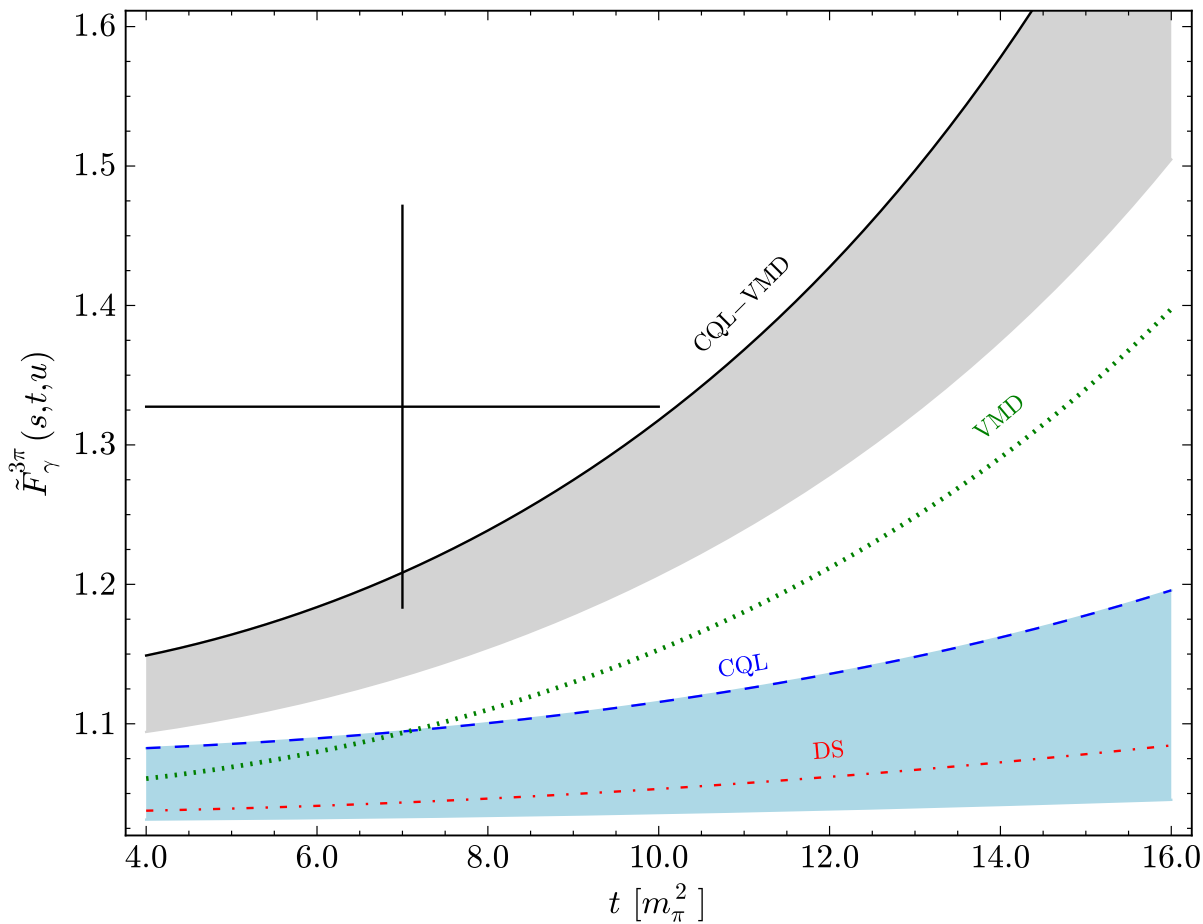}
\caption{(color online) The form factors $\tilde{F}_{\gamma}^{3\pi}(s,t,u)$ 
from various approaches are depicted as functions of $t$, the invariant mass 
of the outgoing pion pair ${\pi^0}(p_2),{\pi^-}(p_3)$, as in the Serpukhov 
and COMPASS experiments, where all pions are on shell.  
We fix $u=m_{\pi}^{2}$ for definiteness.
The upper shaded stripe covers the results of our CQL--VMD approach for  
constituent masses between $M_{q}=360$ MeV (corresponding to the solid 
black curve marking the upper edge of that stripe) and $M_{q}$ equal to 
the DS scale $\Lambda = 565.69$ MeV of Ref. \cite{Bistrovic:1999dy} 
(corresponding to the lower edge of that stripe).
The lower shaded stripe (blue online) covers the results of the 
``pure" CQL model \cite{Bistrovic:1999yy} for the same $M_{q}$ interval.
That is, the (blue) dashed curve depicts the CQL model form factor for 
$M_{q}=360$ MeV, while the lower edge of that stripe is the very slowly 
varying CQL form factor for the high $M_{q} = \Lambda = 565.69$ MeV, 
the DS scale of Ref.  \cite{Bistrovic:1999dy}. A comparison is made with 
results of the ``modified" VMD \cite{Rudaz:1984bz,Holstein:1995qj} (green 
dotted curve) and of DS (in GIA) \cite{Bistrovic:1999dy} (red dash-dotted curve).
Again, the exhibited Serpukhov point \cite{Antipov:1986tp} is actually the 
average value extracted from the total cross-section -- see Subsec. 
\ref{ComparWexperiment} for the comparison with experiment.
}
\label{prim2}
\end{figure}

\begin{figure}[b!]
\includegraphics[scale=0.9]{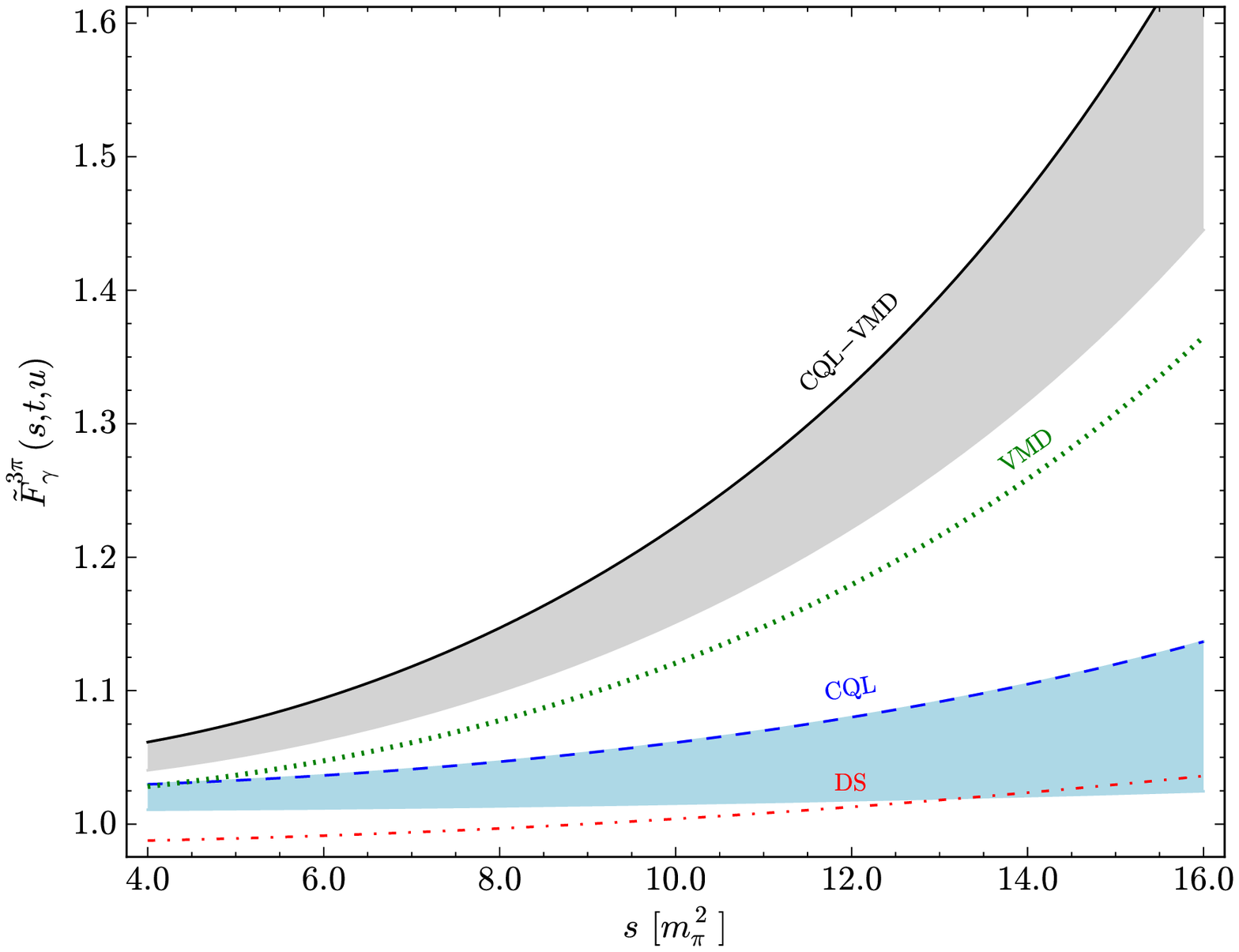}
\caption{(color online) Same as previous Fig. \ref{prim2}, but for the CEBAF kinematics, 
where the two outgoing and on-shell pions are $\pi^{+}$ and $\pi^{0}$. The form factor 
is thus given as the function of their invariant mass squared, the $s$-variable.
For the off-shell pion, $\pi^{-}$, we use $p_{3}^{2}= m_{\pi}^{2}$.
We also fix $t=-m_{\pi}^{2}$. 
}
\label{cebaf2}
\end{figure}

The following Figs. \ref{prim1}--\ref{cebaf2} show the normalized form factor $\widetilde{F}^{3\pi}_{\gamma}=F^{3\pi}_{\gamma}/A^{3\pi}_{\gamma}$. For constituent $u$ and $d$ masses the typical estimates are around $M_{q}\sim m_{p}/3\approx 330$ MeV 
and $M_{q}\sim m_{\rho}/2\approx 385$ MeV, but Figs. \ref{prim1}--\ref{cebaf2}
also show results starting from $M_{q}=300$ MeV and going up to the DS scale 
\cite{Alkofer:1995jx} $\Lambda = 565.69$ MeV. 
The 't Hooft--Veltman integrals, $C_{0}$ and $D_{0}$, were calculated numerically for the case where the photon can be taken on--shell $q^{2}\approx 0$ (pertinent in all experiments \cite{Antipov:1986tp}, \cite{Miskimen94}, \cite{Moinester:1997dm}), and for
\begin{itemize}
\item Primakoff type experiments, \cite{Antipov:1986tp}, \cite{Moinester:1997dm}, 
where all pions are on shell, so that $s+t+u=3 m_{\pi}^{2}$, 
and where we take $u=m_{\pi}^{2}$ for definiteness.
\item CEBAF experiment \cite{Miskimen94}, where the third pion ($\pi^{-}$) is off-shell. The kinematical range explored at CEBAF will be mostly in the $s$ channel, and the amplitudes themselves have a weak dependence on the virtuality of $\pi^{-}$, so we take $p_{3}^{2}\approx m_{\pi}^{2}$. Now $s+t+u=m_{\pi}^{2}$, and we also fix $t=-m_{\pi}^{2}$.
\end{itemize}
Quarks are not confined in our model, so there are possible spurious contributions to the amplitude from the $\bar{q}q$ channel in the box as well as in the triangle if any of the Mandelstam variables $s$, $t$, $u$, is bigger than $4M_{q}^{2}$. 
For $M_{q}$ varied in the range of some $300-500$ MeV, the $q\bar{q}$ thresholds are at $600-1000$ MeV. 
Even the lower value, 600 MeV, is beyond CEBAF upper bound \cite{Miskimen94}: 
$\sqrt{s_{\mathrm{max}}}=4m_{\pi}=554$ MeV (for $m_{\pi}=138.5$ MeV). 
The vector mesons are especially important at COMPASS, where the proposed momentum 
range to be covered should go well above the $\pi^-\pi^0$ threshold, up to the
$\rho$-peak \cite{Kaiser:2008ss}. For such momenta, constituent quark masses 
$M_{q}$ employed in the CQL--VMD approach  should be $M_{q} > m_{\rho}/2$,
say around $400$ MeV, in order to avoid spurious $q\bar{q}$ thresholds.

The general behavior of the amplitudes can be understood through 
three essential factors.
The first is the presence of the  $\rho$ resonance, which is the 
dominant cause of the increase of the amplitude in the present approach.

The second factor is the characteristic mass scale of a given model: in the 
simple CQL (e.g., Ref. \cite{Bistrovic:1999yy}) and the present CQL-VMD models, 
this scale is simply the constituent mass parameter $M_q$, which is typically
not much higher than $\sim\frac{1}{3}M_{\rm nucleon} \sim\frac{1}{2} m_\rho$. 
In DS models, their characteristic scales are also related to their dynamically 
generated momentum-dependent constituent masses and are relatively high compared 
to typical $M_q$ (e.g., $\Lambda=565.69$ MeV in the DS model \cite{Alkofer:1995jx} 
used in Ref. \cite{Bistrovic:1999dy}).
In the CQL model, momentum dependences are stronger for smaller values of 
$M_q$, while for $M_q=\Lambda=565.69$ MeV it is even slightly weaker than 
in the DS approach with this scale $\Lambda$ \cite{Bistrovic:1999dy}. 
(See Figs. \ref{prim1}-\ref{cebaf2} and Ref. \cite{Bistrovic:1999yy}.)
 That larger characteristic mass scales
suppress more the momentum dependence of $F_\gamma^{3\pi}(p_1,p_2,p_3)$, 
is manifest in its power series expansions in CQL \cite{Bistrovic:1999yy}
and DS \cite{Bistrovic:1999dy} papers, where vector mesons were not included.
Nevertheless, it is sufficient to note that the contributions of the quark loops
are suppressed if the quark propagators are suppressed by large masses in their
denominators. Thus, understandably, the present CQL-VMD approach also exhibits 
weaker momentum dependences for for larger $M_q$'s, although now the 
$\rho$-resonance of course dramatically boosts this dependence overall.

The third factor is the symmetry of the $\gamma^* \to 3\pi$ amplitude 
under the interchange of the external momenta $p_i$. It was shown in earlier CQL 
\cite{Bistrovic:1999yy} and DS \cite{Bistrovic:1999dy} approaches without vector 
mesons, most clearly in the aforementioned expansions of $F_\gamma^{3\pi}(p_1,p_2,p_3)$
in powers of the momenta $p_i$ (divided by an appropriate mass scale).
Ref. \cite{Bistrovic:1999dy} clarified how due to this symmetry, the 
contribution of the terms of the second order in momenta [${\cal O}(p^2)$],
is in fact a small constant (of the order of $m_\pi^2$) up to the virtuality of 
the third pion. Therefore, the main contribution, dominating the 
$s,t,u$-dependence for momenta smaller than some characteristic 
model mass scale, comes from ${\cal O}(p^4)$ and not ${\cal O}(p^2)$. This 
gives the parabolic shape to the curves displaying form factors as functions
of Mandelstam variables in Refs. \cite{Bistrovic:1999yy,Bistrovic:1999dy} and
here. In conjunction with the large mass scale $\Lambda$, this also causes 
the weak momentum dependence of $F_\gamma^{3\pi}(p_1,p_2,p_3)$ found 
\cite{Bistrovic:1999dy} in the DS approach using GIA, motivating Ref. 
\cite{Cotanch:2003xv} to go beyond GIA for this process. 

We also show the results for $F_{\gamma}^{3\pi}$ obtained from ``modified" VMD (which is basically our 
Eq. (\ref{cal2}), see also \cite{Rudaz:1984bz}, and \cite{Holstein:1995qj} and references within), 
and DS approach in GIA \cite{Bistrovic:1999dy}.

\subsection{Comparison with experiment}
\label{ComparWexperiment}

We should first note that the Sepukhov experiment \cite{Antipov:1986tp} 
did not, in fact, measure
the presently pertinent form factor ${F}_\gamma^{3\pi}(p_1,p_2,p_3)$, 
but the Primakoff total cross section $\sigma_{tot}$. The latter is thus 
the experimental quantity which is the safest to compare with various 
theoretical predictions, which we do in Fig. \ref{crossSection}.  
The measurements on various targets (with $e Z$ being the nucleus charge) 
yielded $\sigma_{tot}/Z^2 = 1.63 \pm 0.23 \pm 0.13$ nb \cite{Antipov:1986tp},
which is represented by the grey area in Fig. \ref{crossSection}, where it
is compared with the theoretical predictions of VMD and, for various 
constituent quark masses $M_q$ between 300 MeV and 400 MeV, of the CQL and 
CQL--VMD approaches. The former is not compatible with the (admittedly scarce)
experimental data, but its vector-meson extension, the CQL--VMD approach, is.

\begin{figure}[b!]
\includegraphics[scale=0.88]{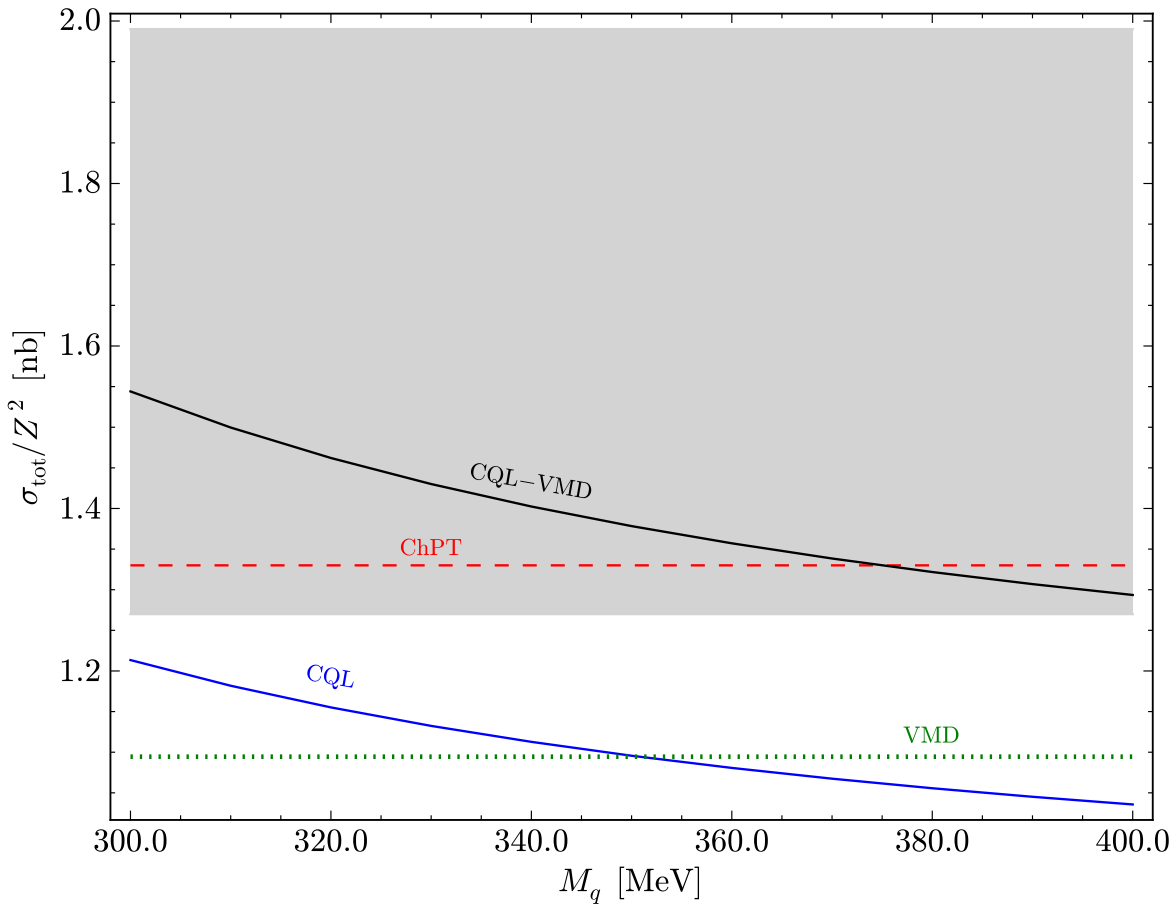}
\caption{(color online) Total cross sections for various $M_q$ predicted by CQM
is depicted by the lower solid curve, while CQM--VMD yielded the upper solid curve.
The (constant) value predicted by VMD alone is depicted by the dotted line, and
the ChPTh prediction (electromagnetically and chiral-loop-corrected) of
$1.33 \pm 0.03$ nb \cite{Ametller:2001yk}, is given by the dashed line.
The experimental cross-section is denoted by the grey area.
}
\label{crossSection}
\end{figure}

This relationship between the CQL and CQL--VMD approaches would seem 
to indicate that agreement with experiment mandates the enhancement due to 
vector mesons and VMD already at the momentum scales where the Serpukhov 
data were gathered.
However, such conclusion would be too rash, since Ametller {\it et al.} 
\cite{Ametller:2001yk} showed that, in this kinematical domain, chiral 
perturbation theory (ChPTh) describes the measured 
$\pi^- \gamma \to \pi^0\pi^-$ process well, after the 
one-photon-exchange electromagnetic corrections are included 
(in their $t$-channel, which is the $u$-channel in our conventions).
This brought their theoretical prediction for the cross-section
to $\sigma_{tot}^{\rm th}/Z^2 = 1.33 \pm 0.03$ nb,
in agreement with experiment.
They also pointed out the importance of the one-loop \cite{Bijnens:1989ff}
and two-loop \cite{Hannah:2001ee} chiral-perturbation-theory contributions,
especially for extracting from the experimental cross section the form factor 
value (call it $F_{0,expt}^{3\pi}=A_{\gamma}^{3\pi}(expt)$) which would correspond 
to the unphysical soft point ($s=t=u=0$). Table I in Ref. \cite{Ametller:2001yk} 
reviews how including 1- and 2-chiral-loop and electromagnetic corrections 
gradually bring about $F_{0,expt}^{3\pi}=10.7\pm 1.2$ GeV$^{-3}$, consistent 
with theory, Eq. (\ref{boxA}). 

These electromagnetic- and loop-corrected ChPTh results indicate that at 
Serpukhov energies VMD is not yet needed for agreement with experiment. 
Thus, the fact that in the COMPASS measurements of $\gamma^* \to 3\pi$, 
the momenta to be covered should surpass the Serpukhov range and approach 
the $\rho$-peak \cite{Kaiser:2008ss} in the vicinity of which VMD dominates,
remains the strongest motivation to combine the simple CQL approach with 
vector mesons and VMD.

\section{Summary}
\label{summary}

The $\gamma^* \to 3\pi$ form factor, presently being measured with high statistics
at CERN by the COMPASS collaboration through the Primakoff experiments 
\cite{Abbon:2007pq,Moinester:1997dm}, has been computed in the
present paper using the simple free constituent quark loop model extended
by vector mesons. This extension turned out to present a problem for the
transitions connecting one vector and three pseudoscalar particles, because
the box graph, which saturates the anomalous amplitude $A_{\gamma}^{3\pi}$
(in the chiral and soft limit), is then supplemented by the $\rho$-resonant
triangle graphs yielding the superfluous contribution of
$\frac{3}{2} A_{\gamma}^{3\pi}$. 
The same problem appears in $\omega\to3\pi$ and $\eta, \eta'\to 2\pi\gamma$ decays, 
where the contribution of the $\rho$-resonant triangles successfully reproduces
the empirical decay widths \cite{Delbourgo:1999qw,Kekez:2005kx}, but the chiral 
and soft--point limit of the pertinent amplitudes are then in conflict with the
low-energy theorems \cite{Aviv:1971hq,Witten:1983tw}.

This problem is cured in the present CQL-VMD approach by removing the 
spurious contribution of the intermediate but non-propagating $\rho$-meson. 
Thanks to this, our
model reproduces correctly the anomalous $\gamma^* \to 3\pi$
chiral- and soft-limit amplitude $A_{\gamma}^{3\pi}$ while including
the effects of the vector mesons at higher momentum scales.

The appropriate subtraction (\ref{corrFF}), of the form of the WT interaction 
\cite{Weinberg:1966zz,Tomozawa:1966jm}, was inspired by the insights
obtained on a more fundamental, microscopic level through DS approach
\cite{Alkofer:1995jx,Bistrovic:1999dy,Maris:1999bh,Cotanch:2003xv}.
The question then arises why not use this, more fundamental QCD-based
approach, to calculate the  $\gamma^* \to 3\pi$  form factor,
instead of the present simplified approach of constituent quarks plus VMD.
In fact, this was done a decade ago in GIA, but the momentum dependence
of the resulting form factor $F_\gamma^{3\pi}(s,t,u)$ is 
then very slow \cite{Bistrovic:1999dy}, for the reasons explained 
in detail in the subsection \ref{completingCalc},  
but also because VMD effects are lacking.
Ref. \cite{Cotanch:2003xv} thus endeavored to reproduce VMD effects working
from a microscopic level, in DS approach, but found that
this problem then requires going beyond GIA, making the task so intractable
in spite of many model simplifications, that only the results for symmetric
kinematics (with at least two Mandelstam variables equal) were given 
\cite{Cotanch:2003xv}. At this point, this DS approach (beyond GIA
and reproducing VMD effects) seems hardly tractable for general kinematics, 
including those of COMPASS and CEBAF. Hence there is a need for related, more
simplified models like the present CQL approach extended by vector mesons,
called the CQL--VMD approach,
where the required features are put in by hand under the guidance
from phenomenology, WI and more microscopic approaches.  Thus we may
consider the CQL--VMD approach as mimicking the more microscopic DS approach
\cite{Cotanch:2003xv} beyond GIA: the relationship between the box graphs
in the respective approaches is obvious, and the (resonant--subtracted)
triangle contributions in the CQL--VMD approach may be regarded as mimicking
the VMD effects reproduced in the DS approach beyond GIA by inserting and
summing up the infinite set of gluon ladder exchanges in the $s, t$ and
$u$--channel of the box graph \cite{Cotanch:2003xv}. However, in contrast
to the very demanding and difficult--to--use DS approach beyond GIA,
the CQL--VMD approach can be easily used for any kinematics that are below the
spurious quark thresholds.

The DS approach beyond GIA, like in Ref. \cite{Cotanch:2003xv} but for 
general kinematics, is a very difficult task which has to be relegated
to a future work. The same holds for calculating electromagnetic corrections,
since in contrast to the corresponding calculation of Ametller et al. 
\cite{Ametller:2001yk}, 
in the present framework the contributing diagrams contain the 
momentum-dependent form-factor $F_\gamma^{3\pi}(s,t,u)$ in the 
$\gamma{3\pi}$ vertex. It makes this task much more difficult,
but it is obviously necessary in order to reach the next level
of refinement in the present approach. 
A much more straightforward future work in this 
direction will include a CQL--VMD calculation of the reaction 
$K^- \gamma^* \to K^- \pi^0$, which can also be measured by the 
COMPASS collaboration \cite{Abbon:2007pq,Moinester:1997dm}.
We will also test the present approach by applying it to numerous 
meson decays currently studied experimentally by WASA at COSY with 
high precision and statistics \cite{Adam:2004ch,Schadmand:2010zz}, 
such as $\eta, \eta' \to 2 \pi \gamma^{(*)}$.

\section*{Acknowledgment}

\noindent D. Kl. and S. B. were supported through the
project No. 119-0982930-1016 of the Ministry of Science,
Education and Sports of Croatia. D. Kl. and S. B. acknowledge
discussions with D. Horvati\'{c} and D. Kekez. The support by
CompStar network is also acknowledged.



\begin{thebibliography}{100}

\bibitem{Adler:1969gk}
  S.~L.~Adler,
  Phys.\ Rev.\  {\bf 177}, 2426 (1969).
  
\bibitem{Bell:1969ts}
  J.~S.~Bell, R.~Jackiw,
  Nuovo Cim.\  {\bf A60 } (1969)  47-61.
  
\bibitem{Adler:1971nq}
  S.~L.~Adler, B.~W.~Lee, S.~B.~Treiman and A.~Zee,
  Phys.\ Rev.\  D {\bf 4}, 3497 (1971).
  
\bibitem{Terent'ev:1971kt}
  M.~V.~Terent'ev,
  Phys.\ Lett.\  B {\bf 38} (1972) 419.

\bibitem{Terentev:1974vr}
  M.~V.~Terentev,
  Usp.\ Fiz.\ Nauk {\bf 112}, 37 (1974)
  [Sov.\ Phys.\ Usp.\  {\bf 17}, 20 (1974)].
  
\bibitem{Aviv:1971hq}
  R.~Aviv and A.~Zee,
  Phys.\ Rev.\  D {\bf 5} (1972) 2372.
  
\bibitem{Antipov:1986tp}
  Y.~M.~Antipov, V.~A.~Batarin, V.~A.~Bezzubov, N.~P.~Budanov, Y.~P.~Gorin, Y.~A.~Gornushkin, S.~P.~Denisov, S.~V.~Klimenko {\it et al.},
  Phys.\ Rev.\  {\bf D36 } (1987)  21.
 

\bibitem{Amendolia:1985bs}
  S.~R.~Amendolia {\it et al.},
  Phys.\ Lett.\  B {\bf 155}, 457 (1985).

\bibitem{Meshcheryakov+al68}
V. A. Meshcheryakov {\it et al.}, Sov. J. Nucl. Phys. \textbf{2}, 87 (1965); Sov. J. Nucl. Phys. \textbf{7}, 100 (1968).

\bibitem{Blokhintseva+al}
Blokhintseva, Grebernnik, et al., ``Investigation of the Mechanism of Inelastic
Pion-nucleon Interaction at 340 MeV". Paper at High-energy Physics Conference, 
Dubna, 1964.

\bibitem{Miskimen94} R. A. Miskimen, K. Wang, A. Yagneswaran (spokesmen),``Study of the Axial Anomaly using the $\gamma\pi^{+}\to\gamma\pi^{+}$ Reaction Near Threshold", letter of intent, CEBAF-experiment 94-015.


\bibitem{Abbon:2007pq}
  P.~Abbon {\it et al.}  [COMPASS Collaboration],
  Nucl.\ Instrum.\ Meth.\  A {\bf 577}, 455 (2007)
  [arXiv:hep-ex/0703049].

\bibitem{Moinester:1997dm}
  M.~A.~Moinester and V.~Steiner,
  arXiv:hep-ex/9801011.

\bibitem{Kekez:1998xr}
  D.~Kekez, B.~Bistrovi\'c, D.~Klabu\v{c}ar,
  Int.\ J.\ Mod.\ Phys.\  {\bf A14}, 161-194 (1999).
  [hep-ph/9809245].

\bibitem{Wess:1971yu}
  J.~Wess, B.~Zumino,
  Phys.\ Lett.\  {\bf B37 } (1971)  95.
  
\bibitem{Witten:1983tw}
  E.~Witten,
  Nucl.\ Phys.\  {\bf B223 } (1983)  422-432.
  
\bibitem{Pisarski:1997bq}
  R.~D.~Pisarski, T.~L.~Trueman and M.~H.~G.~Tytgat,
  Phys.\ Rev.\  D {\bf 56}, 7077 (1997)
  [arXiv:hep-ph/9702362].

\bibitem{Kekez:2005kx}
  D.~Kekez, D.~Klabu\v{c}ar, M.~D.~Scadron,
  Fizika {\bf B14 } (2005)  13-30.
  [hep-ph/0503141].

\bibitem{Steinberger:1949wx}
  J.~Steinberger,
  Phys.\ Rev.\  {\bf 76 } (1949)  1180-1186.

\bibitem{Delbourgo:1999qw}
  R.~Delbourgo, D.~-s.~Liu, M.~D.~Scadron,
  Int.\ J.\ Mod.\ Phys.\  {\bf A14 } (1999)  4331-4346.
  [hep-ph/9905501].

\bibitem{Delbourgo:1993dk}
  R.~Delbourgo, M.~D.~Scadron,
  Mod.\ Phys.\ Lett.\  {\bf A10 } (1995)  251-266.
  [hep-ph/9910242].
  
\bibitem{Hakioglu:1990kg}
  T.~Hakioglu, M.~D.~Scadron,
  Phys.\ Rev.\  {\bf D42 } (1990)  941-944.
  
\bibitem{Hakioglu:1991pn}
  T.~Hakioglu, M.~D.~Scadron,
  Phys.\ Rev.\  {\bf D43 } (1991)  2439-2442.
  
\bibitem{Andrianov:1998kj}
  A.~A.~Andrianov, D.~Espriu, R.~Tarrach,
  Nucl.\ Phys.\  {\bf B533 } (1998)  429-472.
  [hep-ph/9803232].
  

\bibitem{Alkofer:2000wg}
  R.~Alkofer and L.~von Smekal,
  Phys.\ Rept.\  {\bf 353}, 281 (2001)
  [arXiv:hep-ph/0007355].
  
\bibitem{Maris:2003vk}
  P.~Maris and C.~D.~Roberts,
  Int.\ J.\ Mod.\ Phys.\  E {\bf 12}, 297 (2003)
  [arXiv:nucl-th/0301049].
  
\bibitem{Fischer:2006ub}
  C.~S.~Fischer,
  J.\ Phys.\ G {\bf 32} (2006) R253
  [arXiv:hep-ph/0605173].

\bibitem{Alkofer:1995jx}
  R.~Alkofer and C.~D.~Roberts,
  Phys.\ Lett.\  B {\bf 369} (1996) 101
  [arXiv:hep-ph/9510284].

\bibitem{Bistrovic:1999dy}
B.~Bistrovi\'{c} and D.~Klabu\v{c}ar,
Phys.\ Lett.\ B {\bf 478} (2000) 127
[arXiv:hep-ph/9912452].

  
\bibitem{Bistrovic:1999yy}
B.~Bistrovi\'{c} and D.~Klabu\v{c}ar,
Phys.\ Rev.\ D {\bf 61} (2000) 033006
[arXiv:hep-ph/9907515].

\bibitem{Venugopal:1998fq}
  E.~P.~Venugopal and B.~R.~Holstein,
  Phys.\ Rev.\  D {\bf 57}, 4397 (1998)
  [arXiv:hep-ph/9710382].

\bibitem{Holstein:1995qj}
  B.~R.~Holstein,
  Phys.\ Rev.\  {\bf D53 } (1996)  4099-4101.
  [hep-ph/9512338].

\bibitem{O'Connell:1995wf}
  H.~B.~O'Connell, B.~C.~Pearce, A.~W.~Thomas, A.~G.~Williams,
  Prog.\ Part.\ Nucl.\ Phys.\  {\bf 39 } (1997)  201-252.
  [hep-ph/9501251].
  
\bibitem{Rudaz:1974wt}
  S.~Rudaz,
  Phys.\ Rev.\  {\bf D10 } (1974)  3857.
  
\bibitem{Ametller:1983ec}
  L.~Ametller, L.~Bergstrom, A.~Bramon and E.~Masso,
  Nucl.\ Phys.\  B {\bf 228}, 301 (1983).
      
\bibitem{Kawarabayashi:1966kd}
  K.~Kawarabayashi, M.~Suzuki,
  Phys.\ Rev.\ Lett.\  {\bf 16 } (1966)  255.
  
\bibitem{Riazuddin:1966sw}
  Riazuddin, Fayyazuddin,
  Phys.\ Rev.\  {\bf 147 } (1966)  1071-1073.
  
\bibitem{Basdevant:1969rw}
  J.~L.~Basdevant, D.~Bessis and J.~Zinn-Justin,
  Nuovo Cim.\  A {\bf 60}, 185 (1969).
  
\bibitem{Rudaz:1984bz}
  S.~Rudaz,
  Phys.\ Lett.\  {\bf B145 } (1984)  281-284.
  
\bibitem{Cohen:1989es}
  T.~D.~Cohen,
  Phys.\ Lett.\  {\bf B233 } (1989)  467.
  
  
\bibitem{'tHooft:1978xw}
  G.~'t Hooft and M.~J.~G.~Veltman,
  Nucl.\ Phys.\  B {\bf 153} (1979) 365.
  
\bibitem{Ametller:2001yk}
  L.~Ametller, M.~Knecht, P.~Talavera,
  Phys.\ Rev.\  {\bf D64}, 094009 (2001).
  [hep-ph/0107127].


\bibitem{Cotanch:2003xv}
  S.~R.~Cotanch and P.~Maris,
  Phys.\ Rev.\  D {\bf 68} (2003) 036006
  [arXiv:nucl-th/0308008].
  
\bibitem{Maris:1999bh}
  P.~Maris and P.~C.~Tandy,
  Phys.\ Rev.\  C {\bf 61} (2000) 045202
  [arXiv:nucl-th/9910033].
  
\bibitem{Weinberg:1966zz}
  S.~Weinberg,
  Phys.\ Rev.\ Lett.\  {\bf 17 } (1966)  336-340.
  
\bibitem{Tomozawa:1966jm}
  Y.~Tomozawa,
  Nuovo Cim.\  {\bf A46 } (1966)  707-717.
  
\bibitem{Kaiser:2008ss}
  N.~Kaiser and J.~M.~Friedrich,
  Eur.\ Phys.\ J.\  A {\bf 36} (2008) 181
  [arXiv:0803.0995 [nucl-th]].
  
\bibitem{Maris:1999nt}
  P.~Maris, P.~C.~Tandy,
  Phys.\ Rev.\  {\bf C60 } (1999)  055214.
  [nucl-th/9905056].
  
\bibitem{Bijnens:1989ff}
  J.~Bijnens, A.~Bramon and F.~Cornet,
  Phys.\ Lett.\  B {\bf 237}, 488 (1990).

\bibitem{Hannah:2001ee}
  T.~Hannah,
  Nucl.\ Phys.\  B {\bf 593}, 577 (2001)
  [arXiv:hep-ph/0102213].


\bibitem{Adam:2004ch}
  H.~H.~Adam {\it et al.}  [WASA-at-COSY Collaboration],
  arXiv:nucl-ex/0411038.

\bibitem{Schadmand:2010zz}
  S.~Schadmand  [WASA-at-COSY Collaboration],
  AIP Conf.\ Proc.\  {\bf 1322}, 161 (2010).


\end{thebibliography}
\end{document}